# Observation of magnetization reversal and negative magnetization in $Sr_2YbRuO_6$

Ravi P. Singh and C. V. Tomy

*Department of Physics, Indian Institute of Technology Bombay, Mumbai 400 076, INDIA*
*tomy@phy.iitb.ac.in*

## ABSTRACT

*Detailed magnetic properties of the compound $Sr_2YbRuO_6$ are presented here. The compound belongs to the family of double perovskites forming a monoclinic structure. Magnetization measurements reveal clear evidence for two components of magnetic ordering aligned opposite to each other, leading to a magnetization reversal, compensation temperature ($T^* = 34$ K) and negative magnetization at low temperatures and low magnetic fields. Heat capacity measurements corroborate the presence of two components in the magnetic ordering and a noticeable third anomaly at low temperatures (~15 K) which cannot be attributed the Schottky effect. The calculated magnetic entropy is substantially lower than that expected for the ground states of the ordered moments of $Ru^{5+}$ and $Yb^{3+}$, indicating the presence of large crystal field effects and/ or incomplete magnetic ordering and/or magnetic frustrations well above the magnetic ordering. An attempt is made to explain the magnetization reversal within the frameworks of available models.*

## 1. INTRODUCTION

The subject of magnetization reversal in oxides and intermetallic compounds has received considerable attention recently. The magnetization reversal is usually achieved by applying a large magnetic field in a direction opposite to the aligned moments or by changing the temperature in moderate fields. The temperature induced magnetization reversal, which is quite rare, is found to occur in systems having two or more different types of magnetic ions, positioned at different crystallographic sites [1]. A few ferrimagnetic compounds have shown the temperature induced magnetization reversal effect [2] when the two antiferromagnetically coupled magnetic sublattices exhibited different temperature dependence of the magnetization. Interestingly, there are a few other oxides and intermetallic compounds which show temperature induced magnetization reversal such as $LnVO_3$ ($Ln$ = Y, La, Sm, Nd, etc.) [3-8], (Sm,Gd)Al$_2$ [9], $LnCrO_3$ ($Ln$ = Gd, La$_{0.5}$Pr$_{0.5}$) [10-12], etc. The origin of magnetization reversal in the above compounds is found to be entirely different compared to that of the ferrimagnetic compounds. For example, LaVO$_3$ shows magnetization reversal due to the combined effects of Dzyaloshinsky-Moriya (D-M) interaction [13, 14] and a magnetostrictive distortion induced by orbital moments [15] whereas the magnetization reversal in YVO$_3$ is caused by the competition between single ion anisotropy and D-M interaction [5]. At the same time, the observation of magnetization reversal in NdVO$_3$ and SmVO$_3$ is explained on the basis of $N$-type ferrimagnetism arising from the imbalance of the quenching rate of the orbital moments of V$^{3+}$ ions [7,8]. (Sm,Gd)Al$_2$ causes magnetization reversal due to the compensation between the spin and the orbital parts of the ordered moments [9]. In $LnCrO_3$, this effect is attributed to the polarization of the paramagnetic moments of the $Ln$ ions which align opposite to the canted Cr moments [10]. Here we report the magnetization reversal in a new compound Sr$_2$YbRuO$_6$.

Sr$_2$YbRuO$_6$ belongs to the family of double perovskite ruthenates [16] having the general formula Sr$_2Ln$RuO$_6$ ($Ln$ = Y or rare earth). These compounds form in a monoclinic structure belonging to the space group $P2_1/n$ [17]. The structure of these antiferromagnetic insulators can be formed from the perovskite structure of SrRuO$_3$ by replacing the alternate Ru atoms by rare earth atoms [18]. Due to the monoclinic distortion of the perovskite structure, these compounds are known to exhibit interesting magnetic properties at low temperatures due to the canting of the Ru moments resulting from the Dzyaloshinsky-Moria (D-M) interaction among the antiferromagnetically ordered moments. Even though the magnetic ordering is primarily due to the Ru$^{5+}$ moments ($4d^3$, $J$ = 3/2), the rare earth moments also show magnetic ordering in compounds having magnetic rare earths at temperatures close to the ordering of the Ru moments [19]. $^{151}$Eu Mössbauer measurements [20,21] in $Sr_2EuRuO_6$ have indicated the presence of a large exchange field at the rare earth site (~280 kOe) due to the ordered Ru moments. $^{99}$Ru Mössbauer measurements in $Sr_2YRuO_6$ [22] have also shown the presence of a large exchange field (595 kOe) below the magnetic ordering temperature. The presence of this large exchange field is assumed to be responsible for forcing the rare earth moments to order simultaneously with the Ru moments. Doi et al [23] have reported the magnetization for most of the $Sr_2LnRuO_6$ compounds. They have also reported that the zero field-cooled (ZFC) magnetization in $Sr_2YbRuO_6$ was higher than the field-cooled (FC) magnetization. Neutron diffraction measurements at 10 K in this compound [24] have indicated the existence of antiferromagnetic ordering of both the Ru and Yb moments. Since no detailed magnetization studies exist for this compound, we have performed detailed magnetization and heat capacity measurements on this compound. Our magnetization results show clear evidence for a magnetization reversal, resulting in negative magnetization for low fields ($\leq$ 2 kOe) at low temperatures in the FC measurements. Heat capacity measurements show two well defined peaks that can be attributed to the two magnetic orderings along with a prominent anomaly at low temperatures (~15 K) that cannot be fitted to the usual Schottky anomaly. No other member of this double perovskite family of compounds is known to show magnetization reversal/negative magnetization. The results are analyzed within the frameworks of the available models which explain the magnetization reversal.

## 2. EXPERIMENTAL DETAILS

Samples of $Sr_2YbRuO_6$ were prepared in air by the standard solid state reaction method with the starting materials $SrCO_3$, $Yb_2O_3$ and Ru metal powder. The initial mixture was well ground and heated at 960°C for 24 hours. The final sintering of the pelletized powder was carried out at 1285°C for 24 hours after two intermediate heat treatments followed by grindings. The samples were examined by powder x-ray diffraction using an X'pert PRO diffractometer (PANalytical, Holland) having Cu-$K_\alpha$ radiation. The magnetization measurements were carried out as a function of temperature and magnetic field using a vibrating sample magnetometer (Quantum design, USA). The magnetization measurements were carried out in both the zero field-cooled (ZFC) and field-cooled (FC) modes. In ZFC measurements, the sample was cooled in zero applied fields to 5 K, the required magnetic field was applied and the data were then taken while warming. For FC measurements, the sample was cooled from the paramagnetic state to 5 K in an applied field and the data were recorded while heating the sample. The heat capacity measurements using the re-laxation method were performed in a physical property measurement system (Quantum design, USA) in the temperature range 1.8-300 K.

## 3. RESULTS AND DISCUSSION

The x-ray diffraction pattern was refined by the Reitveld analyses using the Fullprof software. The analyses showed that the compound forms in the required phase. The pattern could be indexed to a monoclinic structure with space group $P2_1/n$ (see Fig. 1). The lattice parameters obtained from the analyses are, $a$ = 5.723(2) Å, $b$ = 5.719(2) Å, $c$ = 8.09(3) Å and $\beta$ = 90.2(1)°, which are in good agreement with those reported earlier [24]. Figure 2 (main panel) illustrates the magnetization of $Sr_2YbRuO_6$ as a function of temperature in the ZFC and FC modes measured in an applied field of 50 Oe. No difference between the ZFC and FC magnetization is observed down to ~46 K. The bifurcation between them then starts and the magnetization curves follow entirely different paths below this temperature. In the ZFC mode, the magnetization shows a maximum at ~44 K (see the lower inset of Fig. 2), goes through a minimum at ~39 K and then increases as the temperature is decreased. This minimum occurs with a negative magnetization if the applied fields are small. However, for higher fields (≥ 500 Oe), the ZFC magnetization is entirely positive with the maximum at ~44 K becoming sharper, as shown in the upper inset of Fig. 2. In contrast, the FC magnetization increases below 44 K, goes through a maximum at 39 K, decreases and then passes through a zero value of magnetization ($M = 0$) at the compensation temperature ($T^*$). Below this compensation temperature, the magnetization is negative down to the lowest temperatures (5 K). No hysteretic behaviour was observed whether the FC magnetization was measured while cooling or warming the sample. Figure 3 shows the FC magnetization curves (normalized to the measuring field) for different applied fields. As the applied magnetic field is increased, $T^*$ shifts to lower temperatures and the negative component of the magnetization decreases. Upper inset of Fig. 3 shows the $T^*$ vs $H$ plot. The actual compensation temperature ($T^* = 34$ K) is taken as the extrapolated value at $H = 0$. For fields $H \geq 3$ kOe, one can observe only positive FC magnetization, even though a minimum occurs at temperatures corresponding to $T^*$. Lower inset of Fig. 3 shows additional evidence for the magnetization reversal. For obtaining this data, the sample was field cooled in 10 kOe down to 5 K and the applied field was removed. The remnant magnetization was then measured in zero field while warming the sample. The sudden increase in magnetization before attaining the paramagnetic state can be attributed to the magnetization reversal. Thus the magnetization data (Figs. 2 and 3) clearly show evidence for the magnetic ordering as well as the magnetization reversal/ negative magnetization. In order to verify the reproducibility of the anomalous behaviour, the compound was prepared in two different batches and both showed similar structural and magnetic properties.

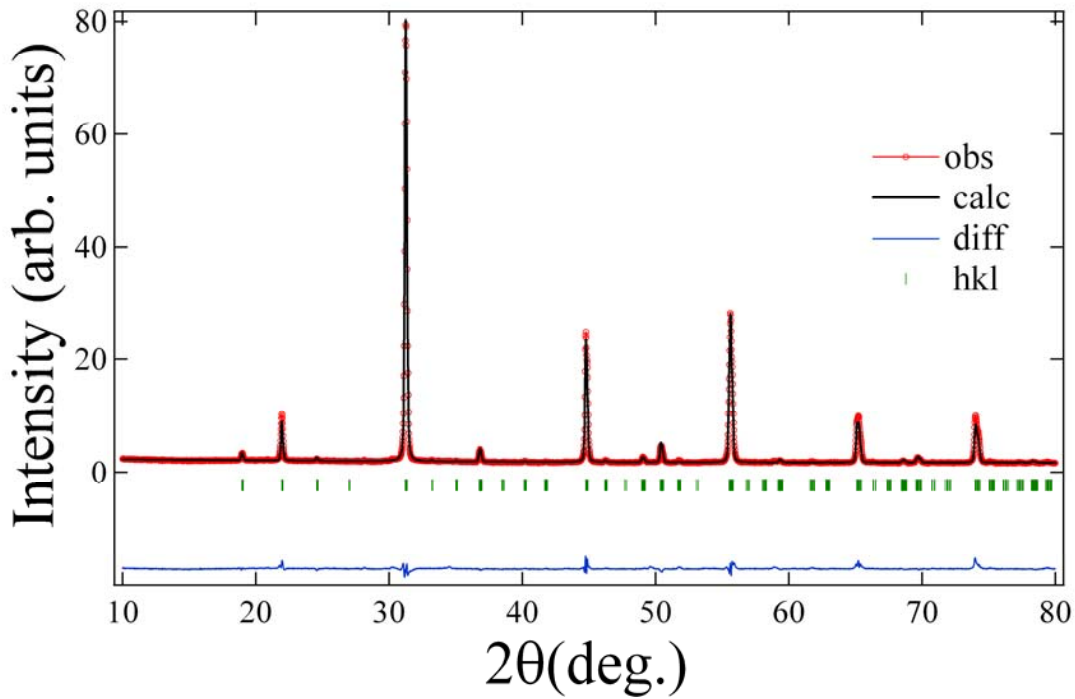

Figure 1. Reitveld analysis of the XRD pattern of $Sr_2YbRuO_6$. The bottom line shows the difference plot between the observed and calculated x-ray intensities.

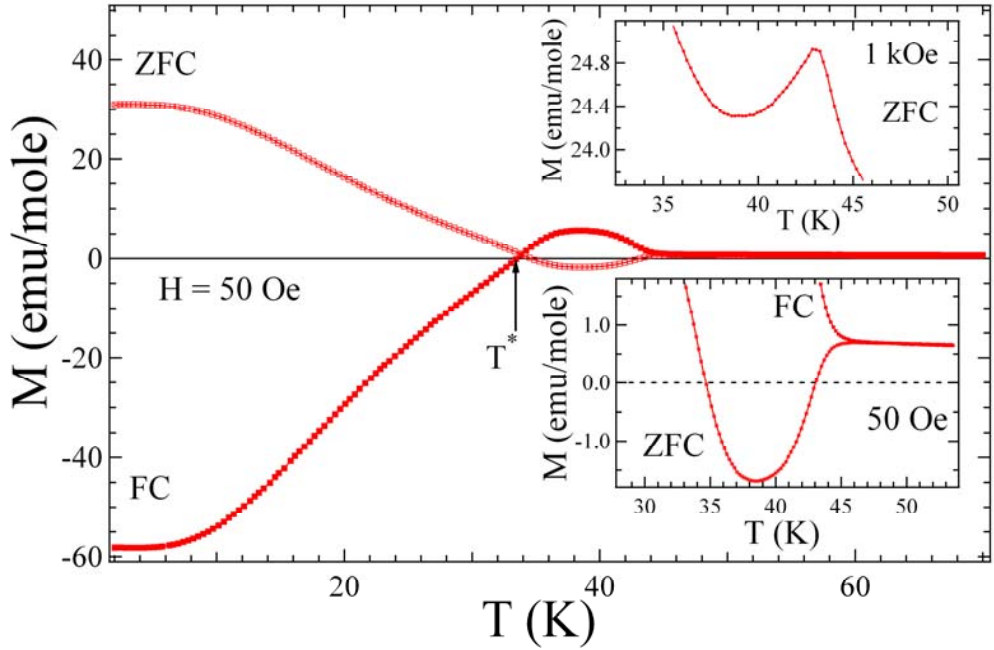

Figure 2. Magnetization of $Sr_2YbRuO_6$ as a function of temperature in the ZFC and FC modes in 50 Oe. Insets show the expanded version of the ZFC mode near the magnetic transitions for 50 Oe (lower) and 1 kOe (upper).

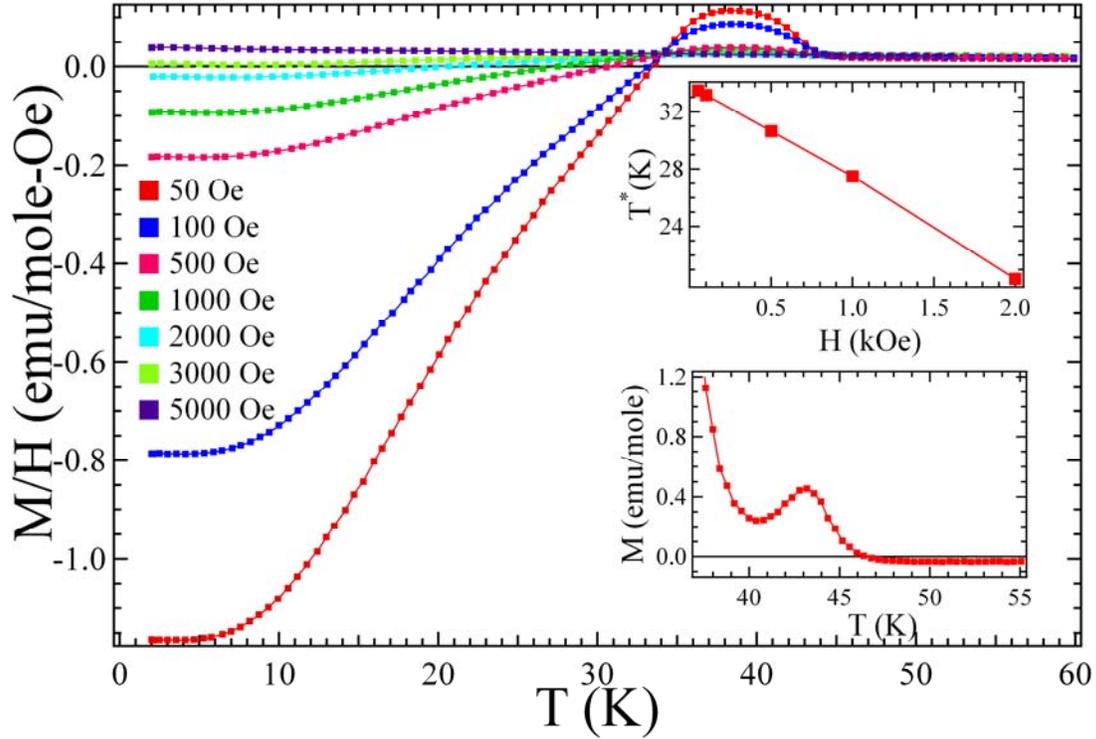

Figure 3. Field-cooled (FC) magnetizations normalized to the measuring fields (*M/H*) as a function of temperature (*T*) at different fields. Upper inset shows the plot of compensation temperature $T^*$ as a function of measuring field *H*. Lower inset shows the warm up data (FCW) in zero field. The sample was field-cooled in 10 kOe magnetic field down to 5 K and the field was switched off before taking the data.

In order to explore the magnetic behaviour in detail, magnetic isotherms were measured at different temperatures between 5 K and 50 K. Figure 4 shows the magnetization curves for some selected temperatures in the ZFC mode. A clear hysteresis is observed at low temperatures in low fields. The magnetization does not saturate even at high fields (90 kOe) and shows only a linear variation with *H* as expected for an antiferromagnet (Fig. 4(h)). As the temperature increases, the hysteresis loop shrinks and the coercive field $H_c$ decreases. A small increase in $H_c$ is observed in the region of 39-44 K (see Fig. 4(e) and (g)). This temperature range corresponds to the onset of the two magnetic anomalies. The magnetic isotherms were also measured in the FC mode which gave similar hysteresis behaviour and $H_c$ variation. Some of the $H_c$ values from the FC loops are also plotted in Fig. 4(g). The above observations clearly indicate the presence of two components in the magnetic ordering, one starting at 44 K and the other at 39 K. The maximum at 44 K and the minimum at 39 K in ZFC magnetization as well as the maximum at 39 K in the FC magnetization correspond to the two components mentioned earlier. From the magnetization behaviour in the FC and ZFC modes, it is clear that the two magnetic components align opposite to each other. Since the magnetic ordering of both the Yb and Ru moments is inferred from the neutron diffraction measurements at 10 K [24], one can attribute the anomalies in the magnetization to the magnetic ordering of Ru and Yb. As the temperature variation of the intensity of the magnetic peaks is not reported in the neutron diffraction measurements [24], it will be difficult to assign the actual ordering temperatures for Yb and Ru. However, since Ru is seen to be ordering first in all the reported compounds of this series [19] and the rare earth moments are forced to order due to the large exchange field [20-22] resulting from the ordered Ru moments, we also assert that the first transition at 44 K is due to the magnetic ordering of the Ru moments.

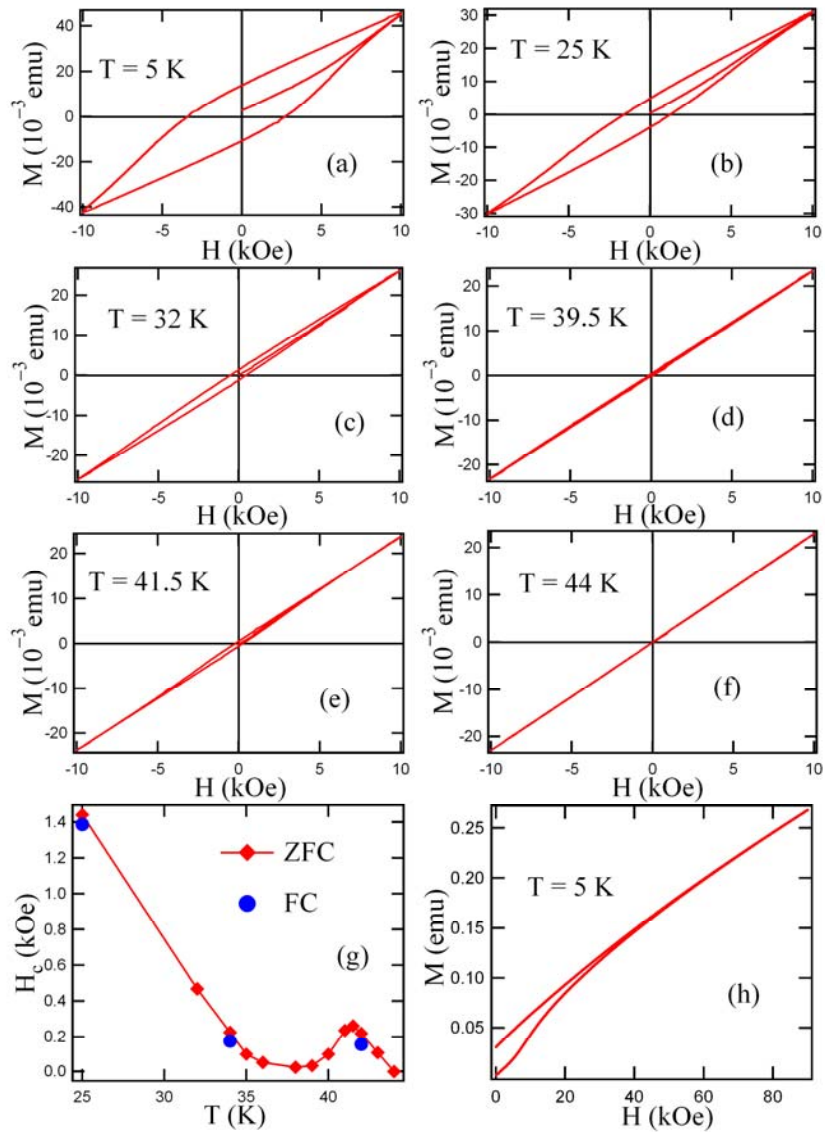

Figure 4. (a) – (f) Magnetization (*M*) as a function of field (*H*) at different temperatures in zero field cooled mode. (g) Coercivity ($H_c$) as a function of temperature. (h) The magnetization extended to high fields at 5 K.

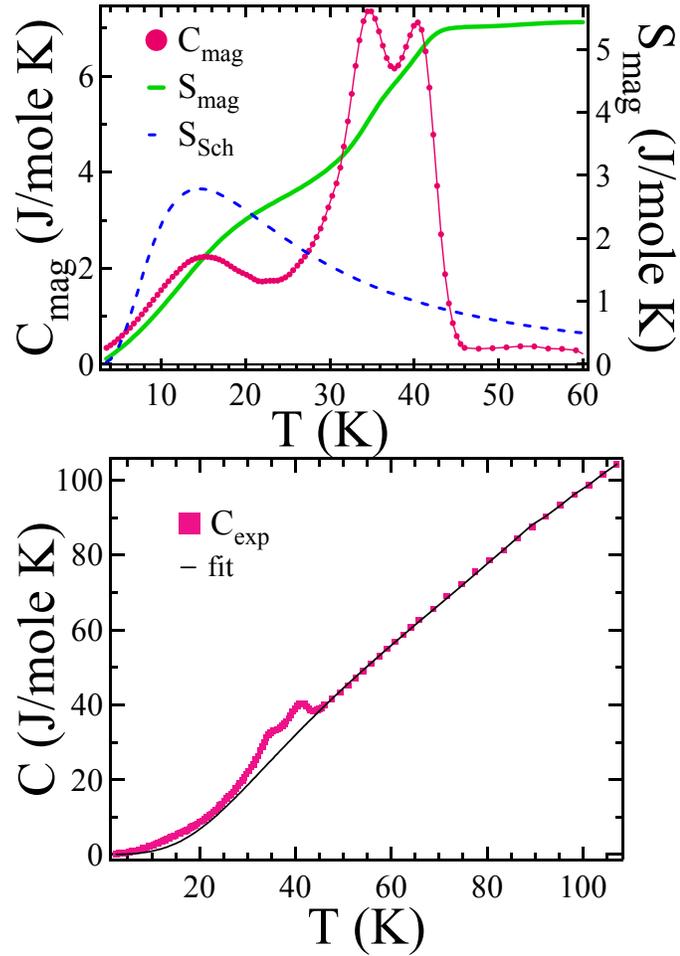

Figure 5. Total heat capacity ($C$) of $Sr_2YbRuO_6$ measured as a function of temperature (lower panel). The solid line represents the phonon contribution calculated using the Debye and Einstein contributions (see text). The upper panel shows the magnetic contribution to the heat capacity ($C_{mag}$) and magnetic entropy ($S_{mag}$). The dashed line is the calculated Schottky contribution to the heat capacity for a simple two-level configuration (ground state and excited state).

There are no reports about the heat capacity of this compound in the literature. The result of heat capacity measurements for $Sr_2YbRuO_6$ is presented in the lower panel of Fig. 5. Two peaks can be identified, one at $T = \sim 44$ K and the second at $T = \sim 39$ K. As inferred from the magnetization measurements, one can attribute the two peaks in heat capacity to the magnetic ordering of the Ru and Yb moments. In order to separate the magnetic contribution to the heat capacity, the phonon contribution needs to be removed from the total measured heat capacity. Since no obvious nonmagnetic analogue is available for this compound, an approximate phonon contribution was calculated using the combined Debye and Einstein terms [25] from the equation,

$$C_{lattice} = R\left\{\sum_{i=1}^{3n-n}\frac{1}{1-\alpha_E T}\frac{(\theta_{E_i}/T)^2 \exp(\theta_{E_i}/T)}{(\exp(\theta_{E_i}/T)-1)^2} + \frac{9}{1-\alpha_D T}\left(\frac{1}{x_D}\right)^3 \int_0^{x_D}\frac{x^4 e^x}{(e^x-1)^2}dx\right\} \quad (1)$$

where $\alpha$'s are the anharmonicity coefficients, $\theta_D$ and $\theta_E$ are the Debye and Einstein temperatures, respectively and $x_D = \theta_D/T$. We have used one Debye and three Einstein frequencies for the calculations along with a single $\alpha_E$. The calculated lattice heat capacity is shown as a solid line in the lower panel of Fig. 5. A reasonably good agreement with the experimental data is observed at high temperatures (above the magnetic ordering). The values of the parameters obtained from the fit are: $\theta_{E1} = 300$ K, $\theta_{E2} = 529$ K, $\theta_{E3} = 700$ K, $\theta_D = 189$ K, $\alpha_E = 1.0 \times 10^{-4}$ K$^{-1}$ and $\alpha_D = 8.4 \times 10^{-4}$ K$^{-1}$. The values obtained for the Debye and Einstein temperatures are comparable with those obtained under similar treatment of the heat capacity for $YVO_3$ which also shows magnetization reversal, compensation and negative magnetization [26]. The contribution of the magnetic heat capacity, $C_{mag}$ was obtained by subtracting the calculated phonon contribution from the total heat capacity and is shown in the upper panel of Fig. 5 along with the magnetic entropy $S_{mag} = \int_{T_1}^{T_2}\frac{C_{mag}}{T}dT$. The magnetic heat capacity $C_{mag}$ clearly shows two peaks one at ~41 K and the second at ~36 K corroborating the presence of two long range orderings as deduced from the magnetization measurements (even though there is a slight mismatch between the corresponding temperatures in the two measurements). The small hump observed at low temperatures (~15 K) was treated as the Schottky anomaly arising from the thermally populated excited levels of Yb (such a peak is absent in the heat capacity of $Sr_2YRuO_6$ [27], isostructural compound with nonmagnetic rare earth) due to the removal of the ground state degeneracy when the crystal field effects are present. First we assumed a simple two level system for Yb energy levels and calculated the total heat capacity by thermal population using the equation

$$C_{Sch} = \frac{R(\Delta/T)^2 (g_0/g_1)\exp(\Delta/T)}{[1+(g_0/g_1)\exp(\Delta/T)]^2} \tag{2}$$

where $R$ is the gas constant, $\Delta$ is the energy difference between the ground state and the excited state (in units of temperature), and $g_0$ ($g_1$) is the multiplicity of the ground (excited) state. For simplicity, we have assumed $g_0 = g_1 = 1$. The Schottky curve obtained using the above equation is shown as the dashed line in the upper panel of Fig. 5 for $\Delta = 35$ K. The large mismatch of the heat capacity values between the calculated and observed curves (the observed anomaly is nearly half the expected value) indicates that the low temperature hump in $C_{mag}$ may not be due to the Schottky anomaly.

Magnetic entropy saturates nearly to a value of ~5.7 J mole$^{-1}$ K$^{-1}$ above 45 K which is less than the entropy expected for the ordered Ru$^{5+}$ moments with a ground state of $J = 3/2$ ($S_{mag} = R\ln(2 \times \frac{3}{2}+1) = 11.52$ J mole$^{-1}$ K$^{-1}$). However, if the crystalline electric field effects are present, then the four-fold degenerate ground state of Ru$^{5+}$ can transform into a Kramer's doublet, giving rise to a multiplicity of only two [24]. This will reduce the magnetic entropy of the compound from 11.52 J mole$^{-1}$ K$^{-1}$ to $R\ln 2 = 5.76$ J mole$^{-1}$ K$^{-1}$. However, the magnetic moment value of Ru$^{5+}$ deduced from the neutron diffraction data at 10 K (3.0 $\mu_B$/Ru$^{5+}$) in Sr$_2$YbRuO$_6$ [24] corresponds well with the expected value of the moment with a ground state $J = 3/2$. The exact reason for such a reduction in the entropy is not very clear at present. If the entropy of the ordered Yb moments is also taken into account (magnetic ordering of Yb$^{3+}$ is inferred from the neutron diffraction data at 10 K [24]), then the discrepancy in entropy is even more serious. A similar entropy reduction was reported in a related compound Ba$_2$YbRuO$_6$ [24]. In this compound also, the calculated entropy was only 7.0 J mole$^{-1}$ K$^{-1}$ even though both the Ru and Yb moments show magnetic ordering at 10 K. After deducting the magnetic entropy observed (3.9 J mole$^{-1}$ K$^{-1}$) for the isostructural compound with nonmagnetic rare earth, Ba$_2$YRuO$_6$, where only the Ru moments show magnetic ordering, the remaining entropy (3.1 J mole$^{-1}$ K$^{-1}$) was attributed to the crystal field split ground state doublet $\Gamma_6$ of Yb$^{3+}$ ions. If we do a similar treatment here and deduct the observed magnetic entropy (2.6 J mole$^{-1}$ K$^{-1}$) of Sr$_2$YRuO$_6$ [27] from the total entropy of Sr$_2$YbRuO$_6$, then we obtain 3.1 J mole$^{-1}$ K$^{-1}$ as the magnetic entropy of Yb$^{3+}$ in our compound also. Even though this value is still smaller than the required value (5.7 J mole$^{-1}$ K$^{-1}$) for the crystal field split doublet ground state $\Gamma_6$ of Yb$^{3+}$ (under the assumption that the crystal field splitting will be almost similar in the distorted cubic structure of Sr$_2$YRuO$_6$), we consider this as the best possibility and attribute the discrepancy (2.66 J mole$^{-1}$ K$^{-1}$) to the non saturation of the ordered magnetic moments of $Yb^{3+}$ (46%). The reduction in entropy of Ru ordering can be compared to the reduction in entropy of the $V^{3+}$ moments in $YVO_3$ that has been attributed to the frustrations of $V^{3+}$ moments at high temperatures (above magnetic ordering) which decreases the effective contribution of the entropy to the magnetic ordering [26]. In a similar manner, we can attribute the frustration effects of the $Ru^{5+}$ moments at high temperatures for the effective reduction of the entropy associated with the Ru ordering in $Sr_2YRuO_6$. In fact, the presence of frustrations among the $Ru^{5+}$ moments has been inferred as the reason for the reduction in $T_N$ in $Sr_2YRuO_6$ even though the compound has a large value of the exchange integral [28]. This will imply that similar frustrations also exist among the Ru moments at high temperatures in $Sr_2YbRuO_6$.

Two distinct magnetic anomalies can be inferred in the compound from both the magnetization and heat capacity measurements, one at ~44 K and the second at ~39 K. The net result of these anomalies is to give rise to a compensation temperature ($T^* = 34$ K) and a negative magnetization in the FC measurements at low fields ($\leq 2$ kOe). This is possible only if one of the components of the magnetization aligns itself against the applied fields. Since the FC magnetization goes through a positive maximum before going through the compensation point, it is clear that the first component of the magnetic ordering at ~44 K aligns parallel to the field and the second component at ~39 K antiparallel to the magnetic field.

The presence of hysteresis in the magnetization data substantiates the presence of a ferromagnetic component in the magnetic ordering. In compounds with low structural symmetry such as $Sr_2LnRuO_6$ (monoclinic structure), weak ferromagnetic interactions can exist among the antiferromagnetically ordered moments due to canting of the spins resulting from the D-M interactions. In fact, the presence of such ferromagnetic component is well documented in compounds of this series with the nonmagnetic rare earth, $Sr_2YRuO_6$ [22, 29] and $Sr_2LuRuO_6$ [23] resulting from the canting of the antiferromagnetically ordered Ru moments. If we consider the same effect to be present in $Sr_2YbRuO_6$ also, then the increase in magnetization below 44 K can be attributed to the effect of the canting of Ru moments. This brings in the possibility that the second magnetic ordering (Yb moments, as we have assumed) also starts with a canting and as a result, a ferromagnetic component. How this component aligns itself against the field uniquely in the Yb compound is not clear now, since no other compounds in this family show negative magnetization. The large value of negative magnetization at low temperatures and small values of $H_c$ in the temperature range of 39-44 K clearly demonstrate that the second canted component (resulting from the Yb moments) is much larger than the first component (from the Ru moments). The possibility of canting is inferred from the presence of a weak (001) magnetic reflection in the neutron dif-fraction measurements in $Sr_2YbRuO_6$ [24]. The absence of the same weak reflection in the neutron diffraction data in $Sr_2TmRuO_6$ [24] and the presence of a ferromagnetic component in the magnetization data clearly affirm the larger magnitude of the ferromagnetic component in $Sr_2YbRuO_6$.

The second possibility that can be considered as the reason for the magnetization reversal is the competing effects of D-M interaction and the single ion anisotropy of Ru moments, as in the case of $YVO_3$ in the temperature range of 75-110 K [5,6]. In this assumption, the magnetic ordering of the Yb moments can be considered to be purely antiferromagnetic without any canted component. If we compare the observation of negative magnetization in $Sr_2YbRuO_6$ with that of $LaVO_3$ compound, then other mechanisms have to be brought in. In $LaVO_3$, the D-M vector rotates against the magnetic field at the structural transition resulting from the first order magnetostrictive distortion [15]. Since the structure of $Sr_2YbRuO_6$ remains the same at 10 K (as deduced from the neutron diffraction measurements [24], the possibility of any structural transition and hence the rotation of the D-M vector against the field can be ruled out. However, whether such a rotation can be initiated by the magnetic ordering of the Yb moment needs further investigation.

Another possibility for negative magnetization is the polarization of the paramagnetic moments in a direction opposite to the direction of the applied magnetic field, as observed in some $LnCrO_3$ compounds [10-12]. However, if we assume the magnetic ordering of Yb moments to take place at ~39 K and the Ru moments at ~39 K as inferred from the peak in the heat capacity measurements (Fig. 5), then this possibility can be ruled out since no other paramagnetic moments exist in the compound. But if we take the discrepancy between the observed magnetic moment of the ordered Yb moments [24] in neutron diffraction measurements at 10 K (0.98 $\mu_B$) to that of the expected value even for the ground state $\Gamma_6$ with the lowest moment value (1.33 $\mu_B$), then we can assign nearly 26% of the Yb moments not to be ordered. This paramagnetic moment, arising from the large fraction of unordered Yb moments, can polarize against the canted field of Ru moments as in the case of $LnCrO_3$ compounds. If we consider this polarization as the cause of magnetization reversal, then the measured magnetization $M$ should follow the equation [10]

$$M = M_{Ru} + C_{Yb}(H_I + H_a)/(T - \theta) \qquad (3)$$

where $M_{Ru}$ is the canted moment of Ru, $H_I$ is the internal field due to the canted Ru moments, $H_a$ is the applied field, $C_{Yb}$ is the Curie constant and $\theta$ is the Weiss constant. The limitation of this analysis is the assumption of $M_{Ru}$ and $H_I$ to be independent of temperature, which is usually true if $T \ll T_N$. In the present case, this may not be true and hence the values obtained will only be an approximation. The solid line in Fig. 6 shows the fit to the magnetization curve ($H$ = 100 Oe) using the above equation. The parameters obtained from the fit are $M_{Ru}$ = 222 emu/mole, $H_I$ = −7690 Oe and $\theta$ = −55 K. These values are comparable to those obtained for GdCrO$_3$ [11] and La$_{0.5}$Pr$_{0.5}$CrO$_3$ [12]. The value of $C_{Yb}$ (= 2.575) used in the analysis was obtained from the free ion Yb$^{3+}$ paramagnetic susceptibility above $T_N$. The goodness of the fit to Eqn. (3) in Fig. 6 indicates the presence of a considerable fraction of Yb moments that are paramagnetic and contributing to the polarization against the canted Ru moments. Under these circumstances, one can attribute the low temperature anomaly in the heat capacity measurements ($T$~15 K) to the completion of the magnetic ordering of the Yb moments. That is, the Yb moments start ordering at ~39 K due to the large internal field from the ordered Ru moments, but complete the ordering at ~15 K. The negative value of the internal field $H_I$ highlights its direction against both the applied field and the canted Ru moments. We could obtain good fits for magnetization curves with other field values also. The inset of Fig. 6 shows the variation of $M_{Ru}$ (lower) and $H_I$ (upper). The variation of these components is in good agreement with the observed magnetization behaviour. The internal field shows an initial increase with the field, but decreases and changes over to positive values consistent with positive magnetization at high fields. The initial increase is consistent with the increase in negative magnetization for low magnetic fields.

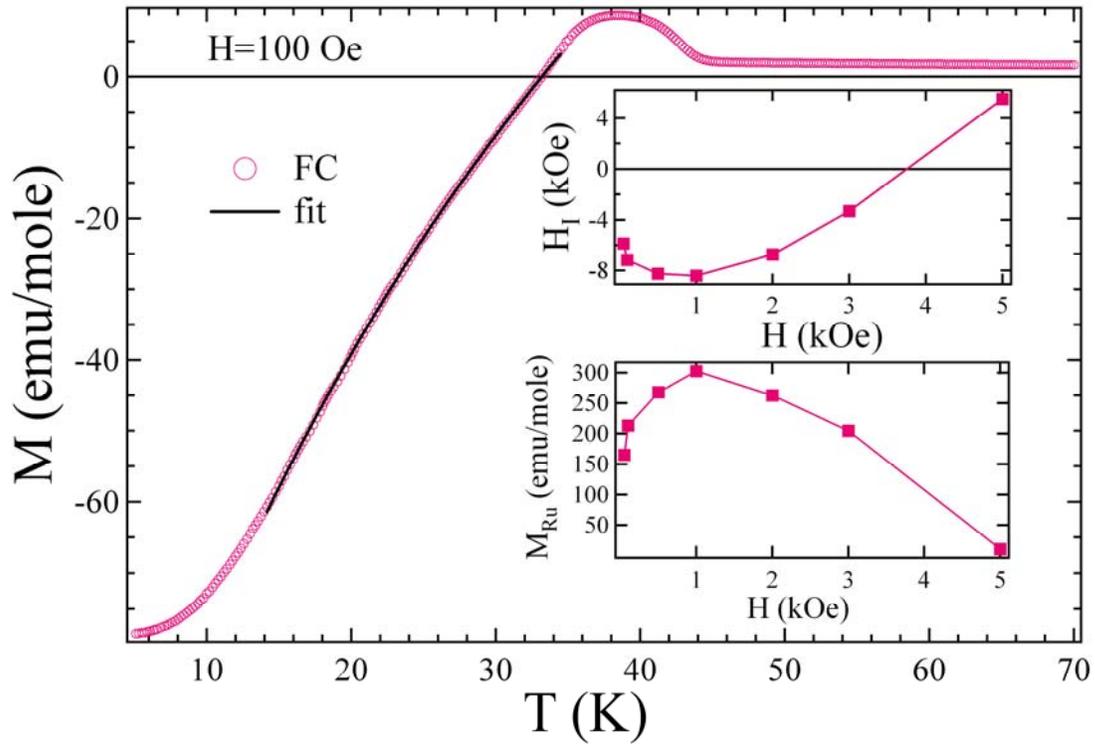

Figure 6. FC magnetization of Sr$_2$YbRuO$_6$ for 100 Oe. The solid line is the fit to the equation (3) in the text. Variation of the canted Ru moment ($M_{Ru}$) (lower panel) and the internal field due to ordered Ru moments ($H_I$) (upper panel) obtained from the fit for different applied fields are given as insets.

In conclusion, we have shown that the antiferromagnetic double perovskite compound $Sr_2YbRuO_6$ shows magnetization reversal, compensation temperature ($M = 0$) and negative magnetization below the magnetic ordering temperatures if the applied magnetic fields are low ($\leq 2$ kOe). Both the magnetization and heat capacity measurements clearly indicate the presence of two components of the magnetic ordering, which are assumed to be due to $Ru^{5+}$ and $Yb^{3+}$ moments. Magnetization reversal is explained using the available models. In order to provide an exact explanation for the observed anomalous behaviour, detailed neutron diffraction measurements are necessary.